\documentclass[twocolumn,pra,aps,amssymb]{revtex4}
\usepackage{amsmath,amssymb,amsfonts,amsthm}
\usepackage{multirow}
\usepackage{verbatim}
\usepackage{url}
\usepackage{comment}
\usepackage{mathtools}
\usepackage{epsf,pgf,graphicx}
\usepackage{color}
\usepackage{tikz,pgf}
\usepackage{gensymb}
\usepackage{algorithm}
\usepackage{color}
\begin{document}
\newcommand{\ket}[1]{\ensuremath{\left|#1\right\rangle}}
\newcommand{\bra}[1]{\ensuremath{\left\langle#1\right|}}
\newcommand\floor[1]{\lfloor#1\rfloor}
\newcommand\ceil[1]{\lceil#1\rceil}
\newtheorem{definition}{Definition}
\newtheorem{theorem}{Theorem}
\newtheorem{claim}{Claim}
\newtheorem{lemma}{Lemma}
\title{Generalized Theoretical Approach for Analysing Optical Experiments}
\author{Arpita Maitra$^1$ and Suvra Sekhar Das$^2$}

\address{$^1$ C R Rao Advanced Institute of Mathematics, Statistics and Computer Science, Hyderabad 500046, India.\\
arpita76b@gmail.com}
\address{$^2$ G.S Sanyal School of Telecommunication, Indian Institute of Technology Kharagpur, Kharagpur 721302, India.\\
suvra@gssst.iitkgp.ac.in} 
\begin{abstract}
A generalized approach towards modelling any optical experiment is presented. Beam splitter and phase retarders are described in terms of annihilation and creation operators. We notice that such description provides us a better way to analyze any optical experiment mathematically than Jones matrix algebra.  We represent polarization of photon in Fock state basis.  We consider recently demonstrated wave-particle superposition generation experiment (Nature Communication, 2017) and Passive BB84 with coherent light (Progress in Informatics, 2011) to test our methodology. We observe that our disciplined methodology can successfully describe the experiments with greater ease, hence offering a convenient tool for modelling any optical arrangement.

\end{abstract}
\maketitle
\section{Introduction}

Enormous proliferation in the domain of Quantum Information, both theoretically and experimentally, leads us to the new era of Quantum Communication. Abstract theoretical models have been implemented in practice. On the other hand, various experimental observations give birth to new theories. However, synthesis of a given experimental setup remains comparatively less explored. For example, experimentally it is shown that one can create a photon which can lie in the superposition state of wave and particle~\cite{waveparticlesuperposition}. Further, entanglement has also been demonstrated between these two behavioural nature of a photon~\cite{waveparticle}.
 Schematic diagram for all these optical setup are available in the literatures. However, those are lacking step by step synthesis of the circuits. 
 
 For involved understanding of a circuit, stepwise synthesis is mandatory. In the present draft, we are searching for a convenient methodology which can serve such purpose. In this direction we exploit second quantization of photon. 

Different kind of Beam Splitters (BS) and various Phase Retarders (PR) are used in any kind of optical experimental setup. We describe different BSs and PRs in terms of annihilation and creation operators. Such description provides us an easier way to analyze any optical setup. One may use Jones matrix algebra to describe any optical device. However, this approach becomes cumbersome when number of photons increase. On the other hand, if we describe the optical devices using second quantization, it becomes handy to model any optical setup for arbitrary number of photons. In this regard, we consider wave-particle superposition~\cite{waveparticle} experiment and Passive BB84 with coherent light~\cite{norbert}. 

We believe that our methodology pave a pathway towards automation, i.e., given an input state and a circuit diagram one may generate an automated algorithm which provides the output result instantly.

 \section{Preliminaries}
In this section we provide the quantum mechanical description of beam splitter and various phase retarders ~\cite{radioeng}. 
\subsection{Beam splitter} 
\begin{figure}[h] 
\begin{center}
\includegraphics[width=0.3\textwidth]{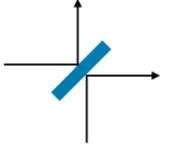}
\caption{Figure shows the schematic diagram of a beam splitter~\cite{agata} (page $2$). Here, incoming ray which is projected on the beam splitter through port $1$ is reflected through port $4$ (vertical arrow) and transmitted through port $3$ (horizontal arrow). The  incoming ray which is projected on the beam splitter through port $2$ is reflected through port $3$ (horizontal arrow) and transmitted through port $4$ (vertical arrow).} 
\label{Fig1}
\end{center}
\end{figure}

The description of quantum Beam Splitter (BS) is available in many literatures~\cite{radioeng,agata,thomas}. The figure (Fig~\ref{Fig1}) of a beam splitter is taken from~\cite{agata} (page $2$). 

\subsubsection{Non Polarizing Beam Splitter}In case of Classical Non Polarizing Beam Splitter (NBS) the input fields are related to the output fields with the following expression.
$$
\begin{pmatrix}
E_3\\
E_4
\end{pmatrix}
=\begin{pmatrix}
t_0 & r_1\\
r_0 & t_1
\end{pmatrix}
.\begin{pmatrix}
E_1\\
E_2
\end{pmatrix}
$$
where $t_0$ (resp. $t_1$) is the transmission coefficient of port $1$ (resp. port $2$) and $r_0$ (resp. $r_1$) is the reflection coefficient of port $1$ (resp. port $2$).

Now, the relation between $r_0, t_0, r_1, t_1$ can be obtained from the following formula~\cite{radioeng}.
$$\arg(r_0)+\arg(r_1)-\arg(t_0)-\arg(t_1)=\pm\pi$$ 
Based on the construction of the BS, the sign of the phase has to be determined. The conventional choice for a BS cube is
$\arg(r_0)=\arg(r_1)=\arg(t_0)=0$, but $\arg(t_1)=\pi$~\cite{radioeng}. Thus, for a $50:50$ BS splitter one can write, 
\begin{eqnarray*}
E_3=\frac{E_1+E_2}{\sqrt{2}}
\end{eqnarray*}
and
\begin{eqnarray*}
E_4=\frac{E_1-E_2}{\sqrt{2}}
\end{eqnarray*}
In case of Quantum NBS, the electric fields are replaced by annihilation operator and creation operator, i.e., in case of quantum NBS, we can write 
$$
\begin{pmatrix}
c\\
d
\end{pmatrix}
=\begin{pmatrix}
t_0 & r_1\\
r_0 & t_1
\end{pmatrix}
.\begin{pmatrix}
a\\
b
\end{pmatrix}
$$
where, $c, d$ are annihilation operator at port $3$ and $4$ respectively and $a, b$ are the annihilation operator at the input ports $1$ and $2$ respectively.
Similarly, the same expression can be written for creation operators $a^{\dag}, b^{\dag}, c^{\dag}$ and $d^{\dag}$. 

If we set $r_0=r_1=\sqrt{\eta}$ and $t_0=t_1=\sqrt{1-\eta}$, then with analogy to classical case one may write,
\begin{eqnarray*}
c=\sqrt{1-\eta}a+\sqrt{\eta}b
\end{eqnarray*}
and
\begin{eqnarray*}
d=\sqrt{\eta}a-\sqrt{1-\eta}b
\end{eqnarray*}
Alternatively, we can write
\begin{eqnarray*}
a=\sqrt{1-\eta}c+\sqrt{\eta}d
\end{eqnarray*}
and
\begin{eqnarray*}
b=\sqrt{\eta}c-\sqrt{1-\eta}d
\end{eqnarray*}
Same expression can be drawn for $a^{\dag}$ and $b^{\dag}$. 

If we consider polarization along with the photon number, then an extra index has to be added with the operators, i.e., instead of $a$ (resp. $b$) we use $a_j$ (resp. $b_k$), where $j, k$ each indicates either horizontal or vertical polarization. Similarly, for creation operator instead of $a^{\dag}$ or $b^{\dag}$, we use $a^{\dag}_j$, $b^{\dag}_k$. Hence, from now on we will write, for annihilation operator 
 \begin{eqnarray*}
a_j=\sqrt{1-\eta}c_j+\sqrt{\eta}d_j
\end{eqnarray*}
and
\begin{eqnarray*}
b_k=\sqrt{\eta}c_k-\sqrt{1-\eta}d_k
\end{eqnarray*}
And for creation operator, we will write
\begin{eqnarray*}
a_j^{\dag}=\sqrt{1-\eta}c_j^{\dag}+\sqrt{\eta}d_j^{\dag}
\end{eqnarray*}
and
\begin{eqnarray*}
b_k^{\dag}=\sqrt{\eta}c_k^{\dag}-\sqrt{1-\eta}d_k^{\dag}
\end{eqnarray*}
The same expression can be drawn if we consider the following BS operator
\begin{eqnarray*}
\hat{B}=e^{i\theta(a^{\dag}b+b^{\dag}a)}
\end{eqnarray*}
In this notation, the relationship between the output ports and the input ports of a BS is expressed as follows.
$$
\begin{pmatrix}
c\\
d
\end{pmatrix}
={\hat{B}}^{\dag}
\begin{pmatrix}
a\\
b
\end{pmatrix}
\hat{B}
$$
Expanding $\hat{B}$ in Taylor series we will get the above expressions for $a$ ($a^{\dag}$) and $b$ ($b^{\dag}$), where $|r_0|=|r_1|=\cos{\theta}$ and $|t_0|=|t_1|=\sin{\theta}$. Here, we assume symmetric BS. In case of $50:50$ BS, $\theta$ will be $\pi/4$.
\subsubsection{Polarizing Beam Splitter} In case of Polarizing Beam Splitter (PBS), Horizontal polarization is transmitted completely where as Vertical polarization is completely reflected. If we assume that the photons which incident on port $1$ transmitted through port $3$ and the photons which incident on port $2$ is transmitted through port $4$, then for Horizontal polarization we can write
\begin{eqnarray*}
a_{H}=c_{H}\\
b_{H}=d_{H}
\end{eqnarray*}
Similarly, if the photon at port $1$ is reflected through port $4$ and the photon at port $2$ is reflected through port $3$, then for Vertical polarization we can write
\begin{eqnarray*}
a_{V}=d_{H}\\
b_{V}=c_{V}
\end{eqnarray*}

\subsection{Phase retarders}\label{pr} 
\begin{figure}[h] 
\begin{center}
\includegraphics[width=0.45\textwidth]{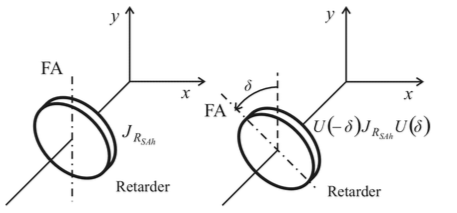}
\caption{Figure shows the schematic diagram of a phase retarder~\cite{radioeng} (page $2$). The left one shows Fast axis parallel to conventional $Y$ axis whereas the right one shows Fast axis making an angle $\delta$ with conventional $Y$ axis.} 
\label{Fig2}
\end{center}
\end{figure}

A schematic diagram of a phase retarder is given in Fig.~\ref{Fig2}. The figure is taken from~\cite{radioeng} (page $2$). 

In case of classical phase retarder (PR), relationship amongst input polarizations and output polarizations can be expressed as 
$$
\begin{pmatrix}
E'_x\\
E'_y
\end{pmatrix}
=\begin{pmatrix}
1 & 0\\
0 & e^{-i\theta}
\end{pmatrix}
.\begin{pmatrix}
E_x\\
E_y
\end{pmatrix}
$$
where, $E'_x$ (resp. $E'_y$) is the output polarization along $X$ (resp. $Y$) axis and $E'_x$ (resp. $E_y$) is the input polarization along $X$ (resp. $Y$) axis~\cite{radioeng}.

In case of quantum PR, we can modify the above expression as
$$
\begin{pmatrix}
a'_x\\
a'_y
\end{pmatrix}
=\begin{pmatrix}
1 & 0\\
0 & e^{-i\theta}
\end{pmatrix}
.\begin{pmatrix}
a_x\\
a_y
\end{pmatrix}
$$
where, $a'_x$ (resp. $a'_y$) is the annihilation operator at output port along $X$ (resp. $Y$) axis and $a_x$ (resp. $a_y$) is the annihilation operator at input port along  $X$ (resp. $Y$) axis.
Thus, we can write
\begin{eqnarray*}
a'_x&=&a_x\\
a'_y&=&e^{-i\theta}a_y
\end{eqnarray*}
where, $\theta$ is the angle made by the PR with its fast axis.

If we assume Horizontal polarization is along $X$ axis and Vertical polarization is along $Y$ axis, then the above equation can be rewritten as
\begin{eqnarray*}
a'_H&=&a_H\\
a'_V&=&e^{-i\theta}a_V
\end{eqnarray*}
\section{Analysis of wave-particle superposition}
In this section we analyze mathematically the experimental setup for generating wave-particle superposition of photon in the light of second quantization of matter. We define two mode polarization basis Fock state $\ket{n_H, n_v}$, where $n_H$ represents the number of horizontally polarized photons and  $n_V$ represents vertically polarized photons with $n_H+n_V=n$ ~\cite{multiphotonrussia}. The basis state can be written in terms of annihilation and creation operator as follows.
\begin{eqnarray*}
\ket{n_H, n_V}=\frac{(a_H^\dag)^{n_{H}} (a_V^\dag)^{n_{V}}}{\sqrt{n_H!n_V!}}\ket{0}.
\end{eqnarray*}
Here, $a^\dag$ stands for creation operator whereas $a$ stands for annihilation operator. An $n$ photon state can be expressed as the superposition of the basis states. That is an $n$ photon state can be written as 
 \begin{eqnarray*}
\ket{\psi^{n}}=\sum_{n_H=0}^ {n} C_{n_H} \ket{n_H, n_V}|_{n_V=n-N_H}
\end{eqnarray*}
where, $\sum_{n_H=0}^{n}|C_{n_H}|^2=1$.

Now, consider the following schematic diagram of the experimental arrangement for generating wave-particle superposition of photon. The diagram is taken from supplementary material of ~\cite{waveparticle} (page $1$ of ~\cite{supp}).
\begin{figure}[h] 
\begin{center}
\includegraphics[width=0.54\textwidth]{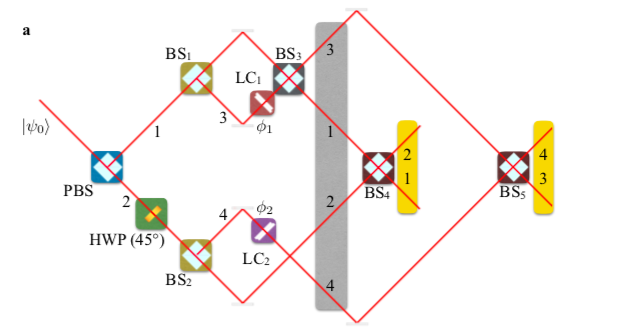}
\caption{Figure shows the layout of the experimental implementation of the wave-particle toolbox (the diagram is taken from the supplementary material of ~\cite{waveparticle}, page $1$~\cite{supp})
}
\label{3}
\end{center}
\end{figure}
The initial photon was prepared in a superposition state of horizontal ($H$) and vertical ($V$) polarization, i.e.,
\begin{eqnarray*}
\ket{\psi^{1}}_{in}=\cos{\alpha}\ket{V}+\sin{\alpha}\ket{H}.
\end{eqnarray*}
In Fock state basis this can be rewritten as 
\begin{eqnarray*}
\ket{\psi^{1}}_{in}&=&\cos{\alpha}\ket{0_H, 1_V}+\sin{\alpha}\ket{1_H, 0_V}\\
&=&(\cos{\alpha}~a_V^{\dag}+\sin{\alpha}~a_H^{\dag})\ket{0_H, 0_V}
\end{eqnarray*}
This photon is then passed through a Polarization Beam Splitter ($PBS$). According to the specification given in ~\cite{waveparticle}, we define
\begin{eqnarray*}
a_H^{\dag}=c_H^{\dag}\\
a_V^{\dag}=d_V^{\dag}\\
b_H^{\dag}=d_H^{\dag}\\
b_V^{\dag}=c_V^{\dag}\\
\end{eqnarray*} 
where $a_H^{\dag}, b_H^{\dag}$ (resp. $a_V^{\dag}, b_V^{\dag}$) represent creation operators at port $1$ and port $2$ for horizontal (Vertical) polarization. Similarly, $c_H^{\dag}, d_H^{\dag}$ (resp. $c_V^{\dag}, d_V^{\dag}$) represent creation operators at port $C$ and port $D$ for horizontal (resp. vertical) polarization.

After $PBS$ the state becomes
\begin{eqnarray*}
\ket{\psi^{1}}_{PBS}=(\cos{\alpha}~d_V^{\dag}+\sin{\alpha}~c_H^{\dag})\ket{0_H, 0_V}.
\end{eqnarray*}
To make the current draft compatible with ~\cite{waveparticle} we define the path of photon which emitted from port $D$ as path $1$ and the path of photon emitted from port $C$ as path $2$. Thus, we rewrite $\ket{\psi^{1}}_{PBS}$ as 
\begin{eqnarray*}
\ket{\psi^{1}}_{PBS}=(\cos{\alpha}~(d_V^{\dag})_1+\sin{\alpha}~(c_H^{\dag})_2)\ket{0_H, 0_V}.
\end{eqnarray*}
The photon in path $1$ is further bifurcated by $50:50$ Beam Splitter ($BS_1$) whereas the photon in pathe $2$ is passed through a Half Wave Plate ($HWP$) making an angle of  $45^{\degree}$ with the fast axis. 
From~\ref{pr} we get that $H$ polarised light passes through the $HWP$ without introducing any phase in the path whereas $V$ polarised light introduce $45^{\degree}$ phase angle into the path. As path $2$ stands for $H$ polarised light, no phase is introduced. Hence, after $BS_1$ and $HWP$ the state becomes

\begin{eqnarray*}
\ket{\psi^{1}}_{BS_1+HWP}&=&(\frac{1}{\sqrt{2}}\cos{\alpha}~((c_V^{\dag})_1+(d_V^{\dag})_3)\\
&+&\sin{\alpha}~(c_H^{\dag})_2)\ket{0_H, 0_V}.
\end{eqnarray*}

Now, photon in path $2$ is bifurcated by $BS_2$. Thus, the resultant state after $BS_1$, $HWP$ and $BS_2$ is written as
\begin{eqnarray*}
\ket{\psi^{1}}_{BS}&=&\frac{1}{\sqrt{2}}(\cos{\alpha}~((c_V^{\dag})_1+(d_V^{\dag})_3)\\
&+&\sin{\alpha}~((c_H^{\dag})_2+(d_H^{\dag})_4))\ket{0_H, 0_V}.
\end{eqnarray*}
Here, path $3$ denotes the photon which is emitted from port $D$ of $BS_1$ and path $4$ represents the photon emitted from port $D$ of $BS_2$. Then both the photon travelling through path $3$ and $4$ are passed through two phase retarders introducing the phase $\phi_1$ in path $3$ and the phase $\phi_2$ in path $4$ respectively. The resultant state now becomes
\begin{eqnarray*}
\ket{\psi^{1}}_{BS}&=&\frac{1}{\sqrt{2}}(\cos{\alpha}~((c_V^{\dag})_1+e^{i\phi_1}(d_V^{\dag})_3)\\
&+&\sin{\alpha}~((c_H^{\dag})_2+e^{i\phi_2}(d_H^{\dag})_4))\ket{0_H, 0_V}.
\end{eqnarray*}
Path $1$ and $3$ now recombined by another $50:50$ Beam Splitter $BS_3$ giving rise to a new state as follows
\begin{eqnarray*}
\ket{\psi^{1}}_{BS}&=&\frac{1}{\sqrt{2}}(\cos{\alpha}~\frac{1}{\sqrt{2}}((c_V^{\dag}+d_V^{\dag})_1+e^{i\phi_1}(c_V^{\dag}-d_V^{\dag})_3)\\
&+&\sin{\alpha}~((c_H^{\dag})_2+e^{i\phi_2}(d_H^{\dag})_4))\ket{0_H, 0_V}\\
&=& \frac{1}{\sqrt{2}}(\cos{\alpha}~\frac{1}{\sqrt{2}}((1+e^{i\phi_1})(c_V^{\dag})_1+(1-e^{i\phi_1})\\
&&(d_V^{\dag})_3)
+\sin{\alpha}~((c_H^{\dag})_2+e^{i\phi_2}(d_H^{\dag})_4))\ket{0_H, 0_V}\\
&=&\frac{1}{\sqrt{2}}(\cos{\alpha}~{\sqrt{2}}e^{\frac{i\phi_1}{2}}(\cos{\frac{\phi_1}{2}}(c_V^{\dag})_1-i\sin{\frac{\phi_1}{2}}(d_V^{\dag})_3)\\
&+&\sin{\alpha}~((c_H^{\dag})_2+e^{i\phi_2}(d_H^{\dag})_4))\ket{0_H, 0_V}\\
\end{eqnarray*}
This can be written as 
\begin{eqnarray*}
\ket{\psi^{1}}_{BS}&=& \cos{\alpha}\ket{wave}+\sin{\alpha}\ket{particle}
\end{eqnarray*}
where, 
\begin{eqnarray*}
\ket{wave}=e^{\frac{i\phi_1}{2}}(\cos{\frac{\phi_1}{2}}(c_V^{\dag})_1-i\sin{\frac{\phi_1}{2}}(d_V^{\dag})_3)
\end{eqnarray*} 
\begin{eqnarray*}
 \ket{particle}=\frac{1}{\sqrt{2}}((c_H^{\dag})_2+e^{i\phi_2}(d_H^{\dag})_4)
 \end{eqnarray*}
 Note that in case of {\em wave} the probability amplitudes of path $1$ and $3$ depend on phase angle $\phi_1$ whereas in case of {\em particle}, the probability amplitudes of path $2$ and $4$ do not depend on phase angle $\phi_2$.
 
 To observe wave-particle morphing as function of $\alpha$, path $1$ and $2$ are  synchronised on beam splitter $BS_4$. Similarly, path $3$ and $4$ are synchronised on beam splitter $BS_5$ resulting the state
 \begin{eqnarray*}
\ket{\psi^{1}}_{BS}&=&\frac{1}{\sqrt{2}}\Big[\Big(\cos{\alpha}e^{\frac{i\phi_1}{2}}\Big(\cos{\frac{\phi_1}{2}}((c_V^{\dag})_1+(d_V^{\dag})_2)\\
&-&i\sin{\frac{\phi_1}{2}}((c_V^{\dag})_3+(d_V^{\dag})_4)\Big)\Big)\\
&+&\sin{\alpha}~\Big(\frac{1}{\sqrt{2}}((c_H^{\dag})_1-(d_H^{\dag})_2)\\
&+&e^{i\phi_2}((c_H^{\dag})_3-(d_H^{\dag})_4)\Big)\Big]\ket{0_H, 0_V}\\
\end{eqnarray*}
where,
\begin{eqnarray*}
\ket{wave-det}&=&e^{\frac{i\phi_1}{2}}\Big(\cos{\frac{\phi_1}{2}}((c_V^{\dag})_1+(d_V^{\dag})_2)\\
&-&i\sin{\frac{\phi_1}{2}}((c_V^{\dag})_3+(d_V^{\dag})_4)\Big)
\end{eqnarray*} 
\begin{eqnarray*}
 \ket{particle-det}&=&\frac{1}{2}e^{i\phi_2}\Big(((c_H^{\dag})_1-(d_H^{\dag})_2)\\
&+&((c_H^{\dag}). _3-(d_H^{\dag})_4)\Big)
  \end{eqnarray*}
   \section{Passive Coherent State BB84~\cite{norbert}}
\begin{figure}[h] 
\begin{center}
\includegraphics[width=0.3\textwidth]{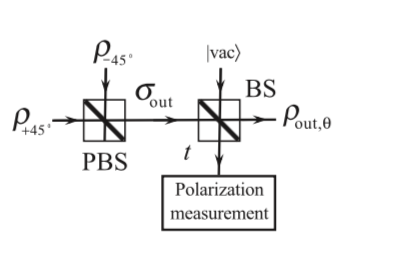}
\caption{Figure shows the schematic diagram of a passive BB84 QKD source with coherent light (page 58 of~\cite{norbert}).} 
\label{fig1}
\end{center}
\end{figure}

   Passive coherent state BB84~\cite{norbert} can be subdivided into two phases. First one is state preparation and the second one in the measurements at Alice and Bob's side. In state preparation phase, Alice starts with two phase randomized strong coherent pluses, prepared in $+45^{\degree}$ and $-45^{\degree}$ linear polarization respectively (Figure~\ref{fig1}). The figure is taken from page number 58 of~\cite{norbert}.
In the figure the state is described in terms of density matrix. In the current draft we consider the vector form of the state, i.e., in our case we will assume that $\ket{\alpha_{+45^{\degree}}}$ is entering from port $1$ of the $PBS$ and $\ket{\alpha_{+45^{\degree}}}$ is entering from port $2$ of the $PBS$. These states can be written as
\begin{eqnarray*}
\ket{\alpha_{\pm45^{\degree}}}&=& 
e^{-\frac{|\alpha|^2}{2}}
\sum_{0}^{\infty}
\frac{\alpha^{n}}{\sqrt{n!}}\ket{n_{\pm 45^{\degree}}}
\end{eqnarray*}
where $n$ is the number of photons in $\pm 45^{\degree}$ angle polarization and $\alpha$ is the mean photon number.

The above expression can be further written as 
\begin{eqnarray*}
\ket{\alpha_{\pm45^{\degree}}}&=& 
e^{-\frac{|\alpha|^2}{2}}
\sum_{0}^{\infty}
\frac{\alpha^{n}}{\sqrt{n!}}(a^{\dag})^{n}\ket{0}
\end{eqnarray*}
which is equivalent to
\begin{eqnarray*}
\ket{\alpha_{\pm45^{\degree}}}&=& 
D(\alpha)\ket{0}\\
&=&e^{\alpha a^{\dag}-\alpha^{*}a}\ket{0}
\end{eqnarray*}
where, $D(\alpha)$ is a unitary operator.

Thus, we can write
\begin{eqnarray*}
\ket{\alpha_{+45^{\degree}}} \ket{\alpha_{-45^{\degree}}}&=& 
D(\alpha_{+45^{\degree}})D(\alpha_{-45^{\degree}})\ket{0}_1\ket{0}_2\\
&=&\exp({\alpha a^{\dag}}_{+45^{\degree}}-{\alpha^{*}a}_{+45^{\degree}})\\
&&\exp({\alpha b^{\dag}}_{-45^{\degree}}-{\alpha^{*}b}_{-45^{\degree}})\ket{0}_1\ket{0}_2
\end{eqnarray*}

This is equivalent to express the state in the following form.
\begin{eqnarray*}
\ket{\alpha_{+45^{\degree}}} \ket{\alpha_{-45^{\degree}}}&=& 
\exp(\frac{\alpha}{\sqrt{2}} (a^{\dag}_{H}+a^{\dag}_{V})
-\frac{\alpha^{*}}{\sqrt{2}}(a_{H}+a_{V}))\\
&&\exp(\frac{\alpha}{\sqrt{2}} (b^{\dag}_{H}-b^{\dag}_{V})
-\frac{\alpha^{*}}{\sqrt{2}}(b_{H}-b_{V}))\\
&&\ket{0}_1\ket{0}_2
\end{eqnarray*}

After Polarization Beam Splitter ($PBS$) the resultant photon states can be written as
\begin{eqnarray*}
\ket{\beta}_{3}\ket{\beta'}_4&=&
\exp(\frac{\alpha}{\sqrt{2}} (c^{\dag}_{H}+d^{\dag}_{V})
-\frac{\alpha^{*}}{\sqrt{2}}(c_{H}+d_{V}))\\
&&\exp(\frac{\alpha}{\sqrt{2}} (d^{\dag}_{H}-c^{\dag}_{V})
-\frac{\alpha^{*}}{\sqrt{2}}(d_{H}-c_{V}))
\ket{0}_3\ket{0}_4
\end{eqnarray*}
Now, rearranging the states we get,
\begin{eqnarray*}
\ket{\beta}_{3}\ket{\beta'}_4&=&
\exp(\frac{\alpha}{\sqrt{2}} (c^{\dag}_{H}-c^{\dag}_{V})
-\frac{\alpha^{*}}{\sqrt{2}}(c_{H}-c_{V}))\\
&&\exp(\frac{\alpha}{\sqrt{2}} (d^{\dag}_{H}+d^{\dag}_{V})
-\frac{\alpha^{*}}{\sqrt{2}}(d_{H}+d_{V}))
\ket{0}_3\ket{0}_4
\end{eqnarray*}
This is equivalent to writing the state as
\begin{eqnarray*}
\ket{\beta}_{3}\ket{\beta'}_4&=&
D(\frac{\alpha}{\sqrt{2}})D(\frac{\alpha}{\sqrt{2}})
\ket{0}_3\ket{0}_4
\end{eqnarray*}
Note that here, $\beta=\beta'=\frac{\alpha}{\sqrt{2}}$.

$\ket{\beta}_3$ is inserted through port $1$ of a Beam Splitter ($BS$) with very low transmission coefficient $t$ ($t<< 1$). In port $2$ of the $BS$ a vacuum state is inserted. Thus we can write,
\begin{eqnarray*}
\ket{\beta}_{1}\ket{0}_2&=&
D(\frac{\alpha}{\sqrt{2}})
\ket{0}_1\ket{0}_2
\end{eqnarray*}
is the input state of $BS$. After $BS$, the state becomes
\begin{eqnarray*}
\ket{\gamma}_3\ket{\gamma'}_4&=&
\exp(\frac{\alpha}{\sqrt{2}}(\sqrt{t}c^{\dag}_{H}+\sqrt{1-t}d^{\dag}_{H}-\sqrt{t}c^{\dag}_{V}-\sqrt{1-t}d^{\dag}_{V})\\
&-&\frac{\alpha^*}{\sqrt{2}}(\sqrt{t}c_{H}+\sqrt{1-t}d_{H}-\sqrt{t}c_{V}-\sqrt{1-t}d_{V}))\\
&&\ket{0}_3\ket{0}_4
\end{eqnarray*}
Rearranging the state we get
\begin{eqnarray*}
\ket{\gamma}_3\ket{\gamma'}_4&=&
\exp(\frac{\alpha\sqrt{t}}{\sqrt{2}}(c^{\dag}_{H}-c^{\dag}_{V})-\frac{\alpha^{*}\sqrt{t}}{2}(c_{H}-c_{V}))\\
&&\exp(\frac{\alpha\sqrt{1-t}}{\sqrt{2}}(d^{\dag}_{H}-d^{\dag}_{V})-\frac{\alpha^{*}\sqrt{1-t}}{\sqrt{2}}(d_{H}-d_{V}))\\
&&\ket{0}_3\ket{0}_4\\
&=&D\Big(\frac{\alpha\sqrt{t}}{\sqrt{2}}\Big)D\Big(\frac{\alpha\sqrt{1-t}}{\sqrt{2}}\Big)\ket{0}_3\ket{0}_4
\end{eqnarray*}
where, $\gamma=\frac{\alpha\sqrt{t}}{\sqrt{2}}$ and $\gamma'=\frac{\alpha\sqrt{1-t}}{\sqrt{2}}$.
Now, $\ket{\gamma}$ is sent to Bob where as $\ket{\gamma'}$ is retained by Alice for polarization measurement.

\begin{figure}[h] 
\begin{center}
\includegraphics[width=0.3\textwidth]{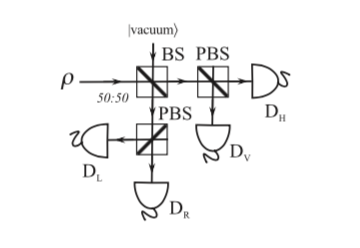}
\caption{Figure shows the schematic diagram of the polarization measurement in a passive BB84 QKD  with coherent light (page 59 of~\cite{norbert}).} 
\label{fig2}
\end{center}
\end{figure}

The measurement setup at Alice's place is described in figure~\ref{fig2}. The figure is taken from page 59 of~\cite{norbert}. 
 
 In our case, $\rho$ in the figure describes the density matrix of $\ket{\gamma'}$. According to the figure, the initial state before $50:50$ BS is 
 \begin{equation}
 \label{equation}
\ket{\gamma'}_1\ket{0}_2=
D\Big(\frac{\alpha\sqrt{1-t}}{\sqrt{2}}\Big)\ket{0}_1\ket{0}_2
\end{equation}
 
 Now, let the annihilation and creation operator at port $1$ of BS be $a$ and $a^{\dag}$ respectively. Similarly, let the annihilation and creation operator at port $2$ of BS be $b$ and $b^{\dag}$ respectively. Then equation~\ref{equation} can be rewritten as
  \begin{eqnarray*}
\ket{\gamma'}_1\ket{0}_2&=&
\exp(\frac{\alpha\sqrt{1-t}}{\sqrt{2}}(a^{\dag}_{H}-a^{\dag}_{V})-\\
&&\frac{\alpha^{*}\sqrt{1-t}}{\sqrt{2}}(a_{H}-a_{V}))
\ket{0}_1\ket{0}_2
\end{eqnarray*}
After beam splitter, the state becomes
 \begin{eqnarray*}
\ket{\delta}_3\ket{\delta'}_4&=&
\exp(\frac{\alpha\sqrt{1-t}}{2}(c^{\dag}_{H}+d^{\dag}_{H}-c^{\dag}_{V}-d^{\dag}_V)-\\
&&\frac{\alpha^{*}\sqrt{1-t}}{2}(c_{H}+d_{H}-c_{V}-d_{V}))
\ket{0}_3\ket{0}_4
\end{eqnarray*}
Rearranging the state we get
\begin{eqnarray*}
\ket{\delta}_3\ket{\delta'}_4&=&
\exp(\frac{\alpha\sqrt{1-t}}{2}(c^{\dag}_{H}-c^{\dag}_{V})-\frac{\alpha^{*}\sqrt{1-t}}{2}\\
&&(c_{H}-c_{V}))
\exp(\frac{\alpha\sqrt{1-t}}{2}(d^{\dag}_{H}-d^{\dag}_{V})-\\
&&\frac{\alpha^{*}\sqrt{1-t}}{2}
(d_{H}-d_{V}))
\ket{0}_3\ket{0}_4
\end{eqnarray*}
According to the figure~\ref{fig2}, $\ket{\delta}$ enters in PBS associated with $\{\ket{H},\ket{V}\}$ basis, whereas $\ket{\delta'}$ enters in PBS associated with $\{\ket{R},\ket{L}\}$ basis. Thus, the initial states
of  the PBS associated with $\{\ket{H},\ket{V}\}$ basis will be
\begin{eqnarray*}
\ket{\delta}_1\ket{0}_2&=&
\exp(\frac{\alpha\sqrt{1-t}}{2}(a^{\dag}_{H}-a^{\dag}_{V})-\frac{\alpha^{*}\sqrt{1-t}}{2}\\
&&(a_{H}-a_{V}))
\ket{0}_1\ket{0}_2. 
\end{eqnarray*}
And the out-coming state from the PBS associated with $\{\ket{H},\ket{V}\}$ basis will be
\begin{eqnarray*}
\ket{\mu}_3\ket{\mu'}_4&=&
\exp(\frac{\alpha\sqrt{1-t}}{2}(c^{\dag}_{H}-d^{\dag}_{V})-\frac{\alpha^{*}\sqrt{1-t}}{2}\\
&&(c_{H}-d_{V}))
\ket{0}_3\ket{0}_4. 
\end{eqnarray*}
Rearrarnging the state we can write,
\begin{eqnarray*}
\ket{\mu}_3\ket{\mu'}_4&=&
\exp(\frac{\alpha\sqrt{1-t}}{2}c^{\dag}_{H}-\frac{\alpha^{*}\sqrt{1-t}}{2}
c_{H})\\
&&\exp(-\frac{\alpha\sqrt{1-t}}{2}d^{\dag}_{V}+\frac{\alpha^{*}\sqrt{1-t}}{2} d_{V})
\ket{0}_3\ket{0}_4. 
\end{eqnarray*}
This implies that at port $3$ of the PBS, we get horizontally polarized coherent state with mean photon number $\frac{\alpha\sqrt{1-t}}{2}$. And at port $4$, we get vertically polarized coherent state with mean photon number $\frac{\alpha^*\sqrt{1-t}}{2}$.

Proceeding in the similar way, one may write the out-coming states from the PBS associated with $\{\ket{R},\ket{L}\}$ basis as 
\begin{eqnarray*}
\ket{\nu}_3\ket{\nu'}_4&=&
\exp(\frac{\alpha\sqrt{1-t}}{2}c^{\dag}_{H}-\frac{\alpha^{*}\sqrt{1-t}}{2}
c_{H})\\
&&\exp(-\frac{\alpha\sqrt{1-t}}{2}d^{\dag}_{V}+\frac{\alpha^{*}\sqrt{1-t}}{2} d_{V})
\ket{0}_3\ket{0}_4. 
\end{eqnarray*}
Describibg in $\{\ket{R}, \ket{L}\}$ basis, one gets
\begin{eqnarray*}
\ket{\nu}_3\ket{\nu'}_4&=&
\exp(\frac{\alpha\sqrt{1-t}}{2\sqrt{2}}(c^{\dag}_{R}+c^{\dag}_{L})-\frac{\alpha^{*}\sqrt{1-t}}{2\sqrt{2}}
(c_{R}+c_{L}))\\
&&\exp(-\frac{\alpha\sqrt{1-t}}{2\sqrt{2}}(d^{\dag}_{R}-d^{\dag}_{L})+\frac{\alpha^{*}\sqrt{1-t}}{2\sqrt{2}} (d_{R}-d_{L}))\\
&&\ket{0}_3\ket{0}_4. 
\end{eqnarray*}
This implies that at each of the output ports of the PBS, we get a linear combination of $\ket{L}$ and $\ket{R}$. Now, we set a detector which detects $\ket{R}$ at port $3$ and a detector which detects $\ket{L}$ at port $4$. So, at port $3$, we can only detect the coherent light with $R$ polarization with mean photon number $\frac{\alpha\sqrt{1-t}}{2\sqrt{2}}$. Similarly, at port $4$ we can only detect the coherent light with $L$ polarization with mean photon number $\frac{\alpha^*\sqrt{1-t}}{2\sqrt{2}}$.
\section{Discussions and Conclusion}
In this section we summarize our disciplined methodology in form of an algorithm. The algorithm is described below.
\begin{enumerate}
 \item Inputs: initial photon state, circuit diagram.
 
 \item Represent the initial photon state in Fock state basis, i.e., in terms of 
 \begin{eqnarray}
\ket{n_H, n_V}=\frac{(a_H^\dag)^{n_{H}} (a_V^\dag)^{n_{V}}}{\sqrt{n_H!n_V!}}\ket{0}.
\end{eqnarray} 
 
 \item If the photon passes through a BS, then for 
 \begin{itemize}
 \item port $1$ and Horizontal polarization $H$, write $a_H=\sqrt{1-\eta}c_H+\sqrt{\eta}d_H$, where $\eta$ is reflection coefficient and $c_{H}$ and $d_{H}$ represent outer ports  of the BS.
 \item  port $2$ and Horizontal polarization $H$, write $b_H=\sqrt{\eta}c_H-\sqrt{1-\eta}d_H$.
 \item  port $1$ and Vertical polarization $V$, write $a_V=\sqrt{1-\eta}c_V+\sqrt{\eta}d_V$, where  $c_{V}$ and $d_{V}$ represent outer ports  of the BS.
 \item port $2$ and Vertical polarization $V$, write $b_V=\sqrt{\eta}c_V-\sqrt{1-\eta}d_V$.
 \end{itemize}
 \item If the photon passes through PBS, then for 
 \begin{itemize}
\item $H$ polarization, 
\begin{itemize}
\item write $a_H=c_H$ 
\item write $b_H=d_H$
\end{itemize}
\item $V$ polarization,
\begin{itemize}
\item write $a_V=d_V$
\item write $b_V=c_V$
\end{itemize}
\end{itemize}
 \item If the photon passes through a PR making an angle $\theta$ with its fast axis, then for 
 \begin{itemize}
\item $H$ polarization, write $a_H=c_H$
\item $V$ polarization write $a_V=e^{-i\theta}c_V$, \\
where $a$ stands for input port and $c$ stands for output port.
\end{itemize}
\item Output; resultant state
 \end{enumerate}
Our methodology may open up an avenue for automation where given an optical circuit and an initial state, the generated output will combine all optical operations that the photon passes through. It may also help for Hamiltonian formulation of Linear Optics ~\cite{Sekiguchi}.

\end{document}